%% file: FieldPredict.tex
\documentclass[oldversion]{aa} 
\usepackage{graphicx}
\usepackage{txfonts}
\usepackage{natbib}
\bibpunct{(}{)}{;}{a}{}{,}

\begin{document}

\title{A magnetic field evolution scenario for brown dwarfs\\ and giant planets}

\author{A. Reiners
  \inst{1}\fnmsep\thanks{Emmy Noether Fellow}
  \and
  U. R. Christensen\inst{2}
}

\offprints{A. Reiners}

\institute{Universit\"at G\"ottingen, Institut f\"ur Astrophysik, Friedrich-Hund-Platz 1, 37077 G\"ottingen, Germany\\
  \email{Ansgar.Reiners@phys.uni-goettingen.de}
  \and
  Max Planck Institute for Solar System Research, Max-Planck-Strasse 2, 37191 Katlenburg-Lindau, Germany\\
  \email{christensen@mps.mpg.de}
}

\date{Received ... / Accepted ...}


\abstract{Very little is known about magnetic fields of extrasolar
  planets and brown dwarfs. We use the energy flux scaling law
  presented by Christensen et al. (2009) to calculate the evolution of
  average magnetic fields in extrasolar planets and brown dwarfs under
  the assumption of fast rotation, which is probably the case for most
  of them. We find that massive brown dwarfs of about 70\,M$_{\rm
    Jup}$ can have fields of a few kilo-Gauss during the first few
  hundred Million years.  These fields can grow by a factor of two
  before they weaken after deuterium burning has stopped. Brown dwarfs
  with weak deuterium burning and extrasolar giant planets start with
  magnetic fields between $\sim$100\,G and $\sim$1\,kG at the age of a
  few Myr, depending on their mass.  Their magnetic field weakens
  steadily until after 10\,Gyr it has shrunk by about a factor of 10.
  We use observed X-ray luminosities to estimate the age of the known
  extrasolar giant planets that are more massive than 0.3\,M$_{\rm
    Jup}$ and closer than 20\,pc. Taking into account the age
  estimate, and assuming sun-like wind-properties and radio emission
  processes similar to those at Jupiter, we calculate their radio flux
  and its frequency. The highest radio flux we predict comes out as
  700\,mJy at a frequency around 150\,MHz for $\tau$~Boo\,b, but the
  flux is below 60\,mJy for the rest. Most planets are expected to
  emit radiation between a few Mhz and up to 100\,MHz, well above the
  ionospheric cutoff frequency.}

\keywords{Planets and satellites: magnetic fields -- Stars: activity -- Stars: low-mass, brown dwarfs -- Stars: magnetic fields}

\maketitle
%

\section{Introduction}

The discovery of hundreds of extrasolar planets during the last two
decades has enabled us to investigate the physical properties of these
objects, and to compare them to the known planets of our own solar
system. The discoveries of brown dwarfs, on the other hand, have
opened our view for this interesting class of objects placed between
stars and planets and sharing physical properties with both groups.
Particularly interesting questions are whether planets, brown dwarfs,
and stars share the same physical principles with regard to their
magnetic dynamos, and what typical field strengths must be expected at
objects for which a positive field detection is missing so far (brown
dwarfs and exoplanets).

Radioemissions from magnetized planets result from the interaction of
their magnetospheres with the stellar wind. This accelerates electrons
to several keV, which emit radio waves at the local cyclotron
frequency \citep[e.g.,][]{Zarka92, Zarka98, Farrell99, Zarka07}. A
similar mechanism, the electron cyclotron maser, was identified as the
dominant source of radio emission for a number of very-low mass stars
and brown dwarfs \citep{Hallinan08}. Radio emissions from extrasolar
planets are particularly interesting because they can outshine the
radio emission from a quiet host star \citep[e.g.,][]{Farrell99,
  Griessmeier05}, and may therefore be utilized to discover extrasolar
planets.  So far, none of the searches for radio emission from
extrasolar planets could present a positive detection. This is not too
surprising because currently available facilities are hardly sensitive
enough to observe the expected weak radio flux.

Both the frequency and the total radio flux depend critically on the
magnetic field strength of the planet, the latter primarily because it
controls the cross section of the magnetosphere interacting with the
stellar wind. In addition, the energy flux of the wind, which depends
on the planet's distance from the host star and on stellar activity,
plays an important role for the total radio flux. The spectrum of the
radio emission is expected to show a sharp cutoff at the electron
cyclotron frequency corresponding to the maximum magnetic field
strength close to the planetary surface. Future observations of the
radio spectrum of extrasolar planets can therefore constrain the
field strength rather reliably.

Usually, one assumes that radio emission can be generated at
extrasolar planets in much the same way as they are generated at
Jupiter.  Recent estimates of radio fluxes from known extrasolar
planets were presented by \citet{Farrell99}, \citet{Lazio04},
\citet{Stevens05}, \citet{Griessmeier07a}, and \citet{Jardine08}.  One
of the large uncertainties in the prediction of radio emission is the
magnetic moment of extrasolar planets.  \citet{Griessmeier04}
discussed the different magnetic moment scaling laws that were
available at that time and applied them to estimate the field of
extrasolar planets.  \citet{Christensen06} derived a magnetic field
scaling relation for planets based on dynamos simulations that cover a
broad parameter range.  Recently, \citet{Christensen09} generalized
the scaling law and showed that its predictions agree with
observations for a wide class of rapidly rotating objects, from Earth
and Jupiter to low-mass main sequence (spectral types K and M) and
T~Tauri stars \citep[see also][]{Christensen09b}.  Here, we revisit
the question of magnetic field strength at brown dwarfs and giant
planets and the estimate for the radio flux from extrasolar planets,
using this scaling law, which we believe to be on more solid grounds
both theoretically and observationally than previously suggested
scaling laws.

\section{Magnetic flux estimate}

Here we use this scaling law in the form given by \citet{Reiners09b},
who expressed the magnetic field strength in terms of mass $M$,
luminosity $L$ and radius $R$ (all normalized with solar values):

\begin{equation}
  \label{eq:law}
  B_{\rm dyn} = 4.8 \times \left(\frac{M L^{2}}{R^7}\right)^{1/6} \rm{[kG]},
\end{equation}

where $B_{\rm dyn}$ is the mean magnetic field strength at the surface
of the dynamo.

In its original form \citep{Christensen09}, the scaling relation
connects the strength of the magnetic field to the one-third power of
the energy flux (luminosity divided by surface area). It is
independent of the rotation rate, provided the latter is sufficiently
high. A weak 1/6-power dependence on the mean density and a factor
describing the thermodynamic efficiency of magnetic field generation
also enter into the scaling law.  The latter factor is found to be
close to one for a large class of objects \citep{Christensen09}, and
this value has been used in Eq.  \ref{eq:law}. Aside from this
assumption, the value of the numerical prefactor in this equations is
only based on the results of dynamo simulations.

For massive brown dwarfs and stars the top of the dynamo is close to
or at the surface of the object and the value of $B_{\rm dyn}$ is
directly relevant for observations that relate to the magnetic field
strength. In giant planets, with $M < 13 M_{\rm J}$ (13 Jupiter
masses), the surface of the dynamo region is at some depth, for
example in Jupiter at approximately 83\,\% of the planet's radius and
the field at the surface is somewhat smaller \citep{Christensen09}.
Higher multipole components, which are assumed to make up half of
$B_{\rm dyn}$, drop off rapidly with radius and are neglected for
simplicity. The radius of objects with masses between 0.3 and 70
$M_{\rm J}$ is nearly constant within $\pm$ 20\% of one Jupiter radius
for ages larger than 200 Myr \citep{Burrows01}. In this case the depth
to the top of the dynamo, at the pressure of metallization of hydrogen
in a H-He planet, is approximately inversely proportional to the
mass. We relate the dipole field strength at the equator of a giant
planet, $B_{\rm dip}^{\rm eq}$, to the overall field strength at the
top of the dynamo, $B_{\rm dyn}$, by

\begin{equation}
  \label{eq:correction}
  B_{\rm dip}^{\rm eq} = \frac{B_{dyn}}{2\sqrt{2}} \left(1 - \frac{0.17}{M/{\rm M}_{\rm J}}\right)^3.
\end{equation}

The factor $2\sqrt{2}$ in the denominator results from the assumption
that at the dynamo surface the dipole field strength is half of the
rms field strength and from the fact that the equatorial dipole field
is $1/\sqrt{2}$ of the rms dipole field.

To estimate the field strength of substellar objects, we employ in
Eq.~\ref{eq:law} radii and luminosities from the evolutionary tracks
calculated by \citet{Burrows93, Burrows97} for substellar objects. Our
model predicts a magnetic field of 9\,G for an object of one $M_{\rm
  J}$ at an age of 4.5\,Gyr. This value agrees well with the polar
dipole field strength of Jupiter, which is 8.4\,G
\citep[e.g.,][]{Connerney93}.

\subsection{Rotation of the planets}

We implicitly assume that the objects are rotating above a critical
rotation velocity, so that our scaling law can be applied.  This seems
to be the case in most of the solar-system giant planets
\citep[][Section\,4]{Christensen09b} and in brown dwarfs
\citep{Reiners08}.  Giant planets with semimajor axes less than about
0.1 -- 0.2\,AU are expected to be slowed down to synchronous rotation
by tidal braking on time scales of less than 1\,Gyr \citep{Seager02}.
However, for planets that are very close to their host star the
synchronous rotation rate may still lie above the critical limit for
our scaling law to apply; the value of the latter is somewhat
uncertain. M-dwarfs with rotation periods up to at least 4 days are
generally found to fall into the magnetically saturated regime
\citep{Reiners09a} for which our scaling law applies.  We emphasize
that our results can be considered as upper limits and would be lower
if the objects for some reason would rotate substantially slower.

\section{Radio flux calculation}

We estimate the magnetospheric radio flux from extrasolar giant
planets following the work of \citet{Stevens05}. This means we assume
that the input power into the magnetosphere is proportional to the
total kinetic energy flux of the stellar wind.  \citet{Griessmeier07a}
discussed different mechanisms depending on the source of available
energy. They also calculate radio emission for the case where the
input power is provided by magnetic energy of the interplanetary
field, for unipolar magnetic interaction, and for stellar flares. In
general, stellar activity of planet host stars is relatively weak so
that we concentrate on the kinetic energy flux as source for the power
input.

\subsection{Radio flux scaling law}

In order to calculate the radio flux of a giant extrasolar planet at
the location of the Earth, we relate it to the radio flux of Jupiter
using Equation (14) from \citep{Stevens05} neglecting the dependence
on the wind-velocity (see below):

\begin{equation}
  \label{eq:radioflux}
  P \propto \frac{1}{d^{2}} \left(\frac{\dot{M}_{\star} M_{\rm dip}}{a^{2}}\right)^{2/3},
\end{equation}

with $P$ the radio flux, $d$ the distance to the system,
$\dot{M}_{\star}$ the stellar mass-loss rate, $M_{\rm dip}$ the dipole
moment of the planet, and $a$ the star-planet distance. The dipole
moment is

\begin{equation}
  M_{\rm dip} = B_{\rm dip}^{\rm eq}  R^3,
\end{equation}

with $R$ the planetary radius.  We assume Jupiter-like values for the
stellar wind velocity, $V_W = 400$\,km\,s$^{-1}$. As discussed in
\citet{Stevens05}, the latter may be an overestimate in particular for
close-in planets, because wind velocities are lower closer to the
star. We also note that young stars have higher mass-loss ratios
\citep{Wood05}, but this does not necessarily affect the wind
velocities \citep[see also][who assume a fixed stellar wind
velocity]{Wood02}. Nevertheless, our uncertainties in the wind
velocities are probably on the order of a factor of two, which means
about a factor of $\sim 3$ (0.5\,dex) uncertainty for individual
stars. We refer to \citet{Stevens05} for further details.  In our
approximation, for a given semi-major axis, mass-loss rate and
distance, the radio flux is only a function of the magnetic moment or
the average magnetic field.

The electron cyclotron frequency near the surface in the polar region
is $f_{\rm ce} [{\rm MHz}] = 2.8 B_{\rm dip}^{\rm pol} [{\rm G}]$,
where $B_{\rm dip}^{\rm pol} = 2 B_{\rm dip}^{\rm eq}$ is the polar
dipole field strength.  For Jupiter it corresponds roughly to the
cutoff frequency observed in the radio emission spectrum
\citep[see][]{Zarka92}. The bulk of Jupiter's radio flux occurs in
roughly the frequency range from 0.1~$f_{\rm ce}$ to the cutoff
frequency.

\subsection{Parameter estimates for known host stars}

\subsubsection{Mass-loss rate}
To calculate the mass-loss rate that is required in
Eq.\,(\ref{eq:radioflux}) for host stars of known extrasolar giant
planets (EGPs), we parameterize the stellar mass-loss rates using the
results from \citet{Wood05} and use X-ray luminosities taken from the
NEXXUS
database\footnote{\texttt{http://www.hs.uni-hamburg.de/DE/For/Gal/Xgroup/\\nexxus/nexxus.html}}
\citep{NEXXUS} in analogy with \citet{Stevens05} but with the updated
parameters from \citet{Wood05}
\begin{equation}
  \frac{\dot{M}_{\star}}{{\rm \dot{M}}_\odot} = \left(\frac{R_\star}{{\rm R}_{\odot}}\right)^2 \left(\frac{F_X}{{\rm F}_{X,\odot}}\right)^{1.34}.
\end{equation}

With $F_X = L_X/(4 \pi R^2)$, we get the mass-loss rate as a function
of X-ray luminosity times $(R_{\star}/{\rm R}_{\odot})^{-0.68}$.  The
latter factor introduces an error on the order of $\sim 10$\,\% if the
radius differs from the solar radius by 20\,\%. A radius offset of
50\,\% (i.e. for example between 0.6 and 0.9\,R$_{\odot}$, which is
the difference in radius between a G-type star and an early-M star)
introduces an error of 24\,\% in the mass-loss rate ($\sim 0.1$\,dex),
which means $\sim 15$\,\% in the radio flux $P$.  Compared to the high
variability in X-ray flux and the large systematic uncertainties of
our approach, we can safely ignore this effect and use for the
calculation of the mass-loss rate the equation

\begin{equation}
  \label{eqn:Mdot}
  \frac{\dot{M}_{\star}}{{\rm \dot{M}}_\odot} \approx \left(\frac{L_X}{{\rm L}_{X,\odot}}\right)^{1.34}.
\end{equation}

\subsubsection{Magnetic Moment: Mass and age estimate}
In order to compute the average surface magnetic field of the EGPs, we
need to estimate mass and age for each system. While their masses are
known except for the projection uncertainty ($M\,\sin{i}$), the age is
more difficult to determine. As a rough estimate, we determine the age
from the X-ray activity seen on the host star. X-ray activity is known
to be related to the rotation of the star, or to the Rossby number $Ro
= P/\tau_{\rm conv}$, with $P$ the rotation period and $\tau_{\rm
  conv}$ the convective overturn time \citep[e.g.,][]{Noyes84,
  Pizzolato03}. For the host stars of the EGPs that we discuss in
Section\,\ref{sec:EGPs}, we calculate the convective overturn time
from the $(B-V)$ color according to the relation in \citet{Noyes84},
$B-V$ was taken from the Hipparcos catalogue
\citep{Hipparcos}. \citet{Mamajek08} provides useful relations between
the Rossby number and normalized X-ray luminosity (Eq.\,\ref{eqn:Ro};
their Eq.~2.6), and between rotation period and the age (and the
color) of a star \citep[Eq.\,\ref{eqn:t}; their Eq~2.2, see
also][]{MH08}.

\begin{eqnarray}
  \label{eqn:Ro}
  Ro & = & P/\tau_{\rm conv} = 0.86 - 0.79 ( \log{L_{\rm X}/L_{\rm bol}} + 4.83)\\
  \label{eqn:t}
   t & = & \left( \frac{P}{0.407 (BV - 0.495)^{0.325} } \right)^\frac{1}{0.566}
\end{eqnarray}

A full discussion of the uncertainties in this activity-age
relationship goes far beyond the scope of our paper and can be found,
e.g., in \citet{Soderblom83}, \citet{Barnes07} and \citet{Mamajek08}.
The derived ages have 1$\sigma$ uncertainties on the order of at least
50\,\%, so these values are really mostly indications of the real age
of the planets. For our scaling relations, however, this large an
error does not significantly affect the results; at the typical age of
the stars, the average magnetic field, radio flux, and peak frequency
that we calculate shrink by approximately 20\,\% if we increase the
age by a factor of two. Note that we use the age only for our
calculation of the average magnetic field and not for an estimate of
the mass-loss rate, which we derive directly from the X-ray flux
according to Eq.\,\ref{eqn:Mdot}.

For the calculation of the Rossby number in Eq.\,\ref{eqn:Ro}, we need
the bolometric luminosity. We calculate the bolometric luminosity from
bolometric corrections $BC_V$ and the distance and $V$-magnitudes
given for each planet's host star in \emph{The Extrasolar Planets
  Encyclopaedia}\footnote{\texttt{exoplanet.eu}}.  We determine $BC_V$
from the color $B-V$ following the tabulated values in
\citet{Kenyon95}. In order to correct for the binaries contained in
the sample, we finally fitted the relation between $B-V$ and
bolometric luminosity neglecting obvious outliers.  We use the
luminosities that result from $B-V$ according to the relation

\begin{equation}
  \log{\frac{L_{\rm bol}}{L_{\odot}} = 1.4531 - 2.0395 \, (B-V). }
\end{equation}

\section{Results}

\begin{figure*}
  \centering
  \mbox{\includegraphics[width=\textwidth]{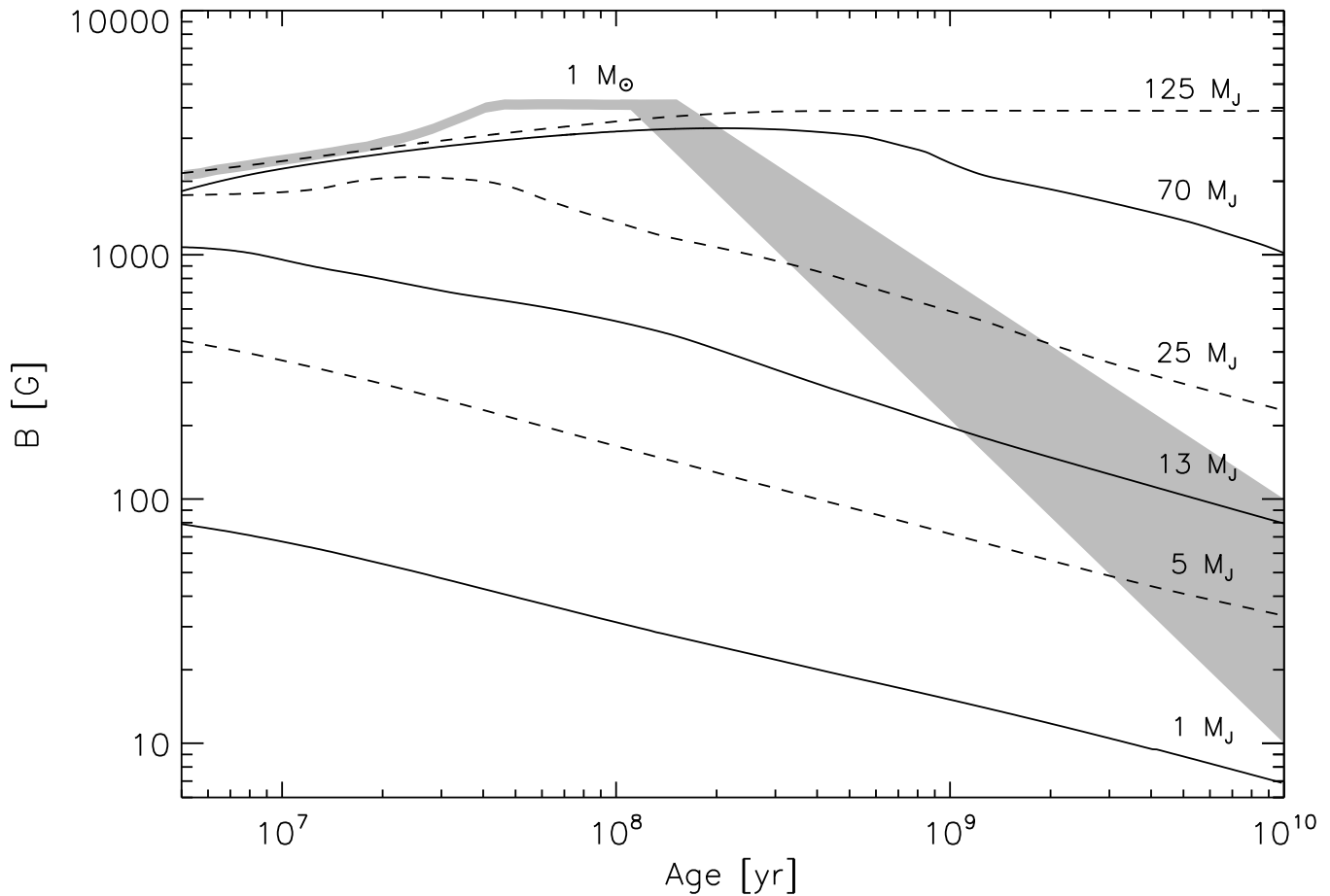}}
  \caption{\label{fig:BfAge}Average magnetic field on the surface of
    the object, $B_{\rm dyn}$, for $M > 13$\,M$_{\rm J}$, and dipole
    field, $B_{\rm dip}^{\rm pol}$, for $M \le 13$\,M$_{\rm J}$, as a
    function of age for giant planets, brown dwarfs, and a very-low
    mass star with $M = $125\,M$_{\textrm{J}}$.  All low-mass objects
    are assumed to be rapidly rotating. An estimate of the average
    magnetic field of the Sun is overplotted \citep[gray shaded area;
    for today's average field see][]{Schrijver87}.}
\end{figure*}

We show the evolution of the magnetic fields in Fig.\,\ref{fig:BfAge}.
For stars and brown dwarfs, i.e., objects with masses higher than
13\,M$_{\rm J}$, we show the average magnetic field of the dynamo,
$B_{\rm dyn}$, which is probably similar to the average surface field.
For giant planets, we plot the dipole field at the surface, $B_{\rm
  dip}$, from Eq.\,\ref{eq:correction}. Evolutionary tracks for a few
planetary mass objects and brown dwarfs are shown, as well as one
model of a very-low mass star (125\,M$_{\textrm J}$). In addition, we
overplot an estimate of the magnetic history of the Sun. The
difference between solar evolution and all other cases considered is
that the Sun developes a radiative core and suffers substantial
rotational braking so that it falls below the saturation threshold
velocity for the dynamo. During the first $\sim 10^8$\,yrs of its
lifetime, the Sun was rotating rapidly enough so that its dynamo
operated in the saturated regime captured by our scaling law.  After
that time, the solar rotation slowed down \citep[e.g.,][]{Skumanich72,
  Barnes07} and magnetic field generation probably weakened in
proportion to angular velocity. Evidence for such a rotation-magnetic
field relation was found in M dwarfs \citep[see][]{Reiners09a}. We
mark this region in grey in Fig.\,\ref{fig:BfAge} because the
uncertainties in the regime of slow rotation are different from our
problem and should not be discussed here. In contrast to sun-like
stars, fully convective stars do not seem to suffer substantial
rotational braking with the result that, even at an age of several
Gyrs, they are rotating in the critical regime for magnetic field
generation \citep{Delfosse98, Barnes07, Reiners08}.

\subsection{Evolution of magnetic fields and radio flux}

The average magnetic fields of giant planets and brown dwarfs
according to our scenario can vary by about an order of magnitude and
more during the lifetime of the objects. The magnetic field strength
is higher when the objects are young and more luminous
(Eq.\,\ref{eq:law}). For example, a one Jupiter-mass planet is
predicted to have a polar dipole field strength on the order of
$100$\,G during the first few Million years; the field weakens over
time and is less than 10\,G after 10\,Gyr. A planet with five Jupiter
masses has a magnetic field at the surface that is consistently
stronger by a factor of four to five over the entire evolutionary
history.

Because of the higher luminosity that is essentially available for
magnetic flux generation, magnetic fields in brown dwarfs are larger
than fields on extrasolar giant planets, varying typically between a
few kG and a hundred G depending on age and mass. Magnetic fields in
brown dwarfs also weaken over time as brown dwarfs cool and loose
luminosity as the power source for magnetic field generation. Low-mass
stars show a generally different behaviour. A low-mass star with $M =
125$\,M$_\textrm{J}$ can produce a magnetic field of about 2\,kG
during the first ten Myr, and the field grows by about a factor of two
until it stays constant from an age of a few hundred Myrs on. For the
solar case, the magnetic field is roughly constant between $5\,10^{7}$
and $10^{8}$\,yrs, which is maintained by the constant luminosity and
rapid rotation.

\subsection{Comparison to other field predictions}

Magnetic field estimates for extrasolar giant planets are available
from a variety of different scaling laws. \citet{Christensen09b}
summarized scaling laws for planetary magnetic fields that were
proposed by different authors. Most of them assume a strong relation
between field strength and rotation rate. As an example,
\citet{Sanchez04} estimated the dipolar magnetic moments of exoplanets
using the ``Elsasser number'' scaling law, which predicts the field to
depend on the square root of the rotation rate but assumes no
dependence on the energy flux. \citet{Sanchez04} predicts average
magnetic fields of $\sim$ 30--60\,G for rapidly rotating planets and
$\sim 1$\,G for slowly rotating ones. The range of values is
comparable to our predictions. If young planets were generally fast
rotators while old planets rotate slowly, which could be the case when
tidal braking plays a role, the results would be similar. However, our
model predicts that energy flux rules the magnetic field strength so
that extrasolar giant planets have high magnetic fields during their
youth and weak magnetic fields at higher ages even if their rotational
evolution is entirely different (given that their are still rotating
fast enough for dynamo saturation).

\citet{Stevens05} used a very simplistic method to scale the magnetic
fields of extrasolar giant planets assuming that the planetary
magnetic moment is proportional to the planetary mass. This implies no
difference between magnetic fields in young and old planets, and no
difference between rapid and slow rotators (but note that slow
rotators were explicitly left out of his analysis). \citet{Stevens05}
also provides radio flux predictions that we compare to our
predictions in the next section.

\subsection{Radio flux and field predictions for known planets}
\label{sec:EGPs}

\begin{figure*}
  \centering
  \mbox{\includegraphics[width=\textwidth]{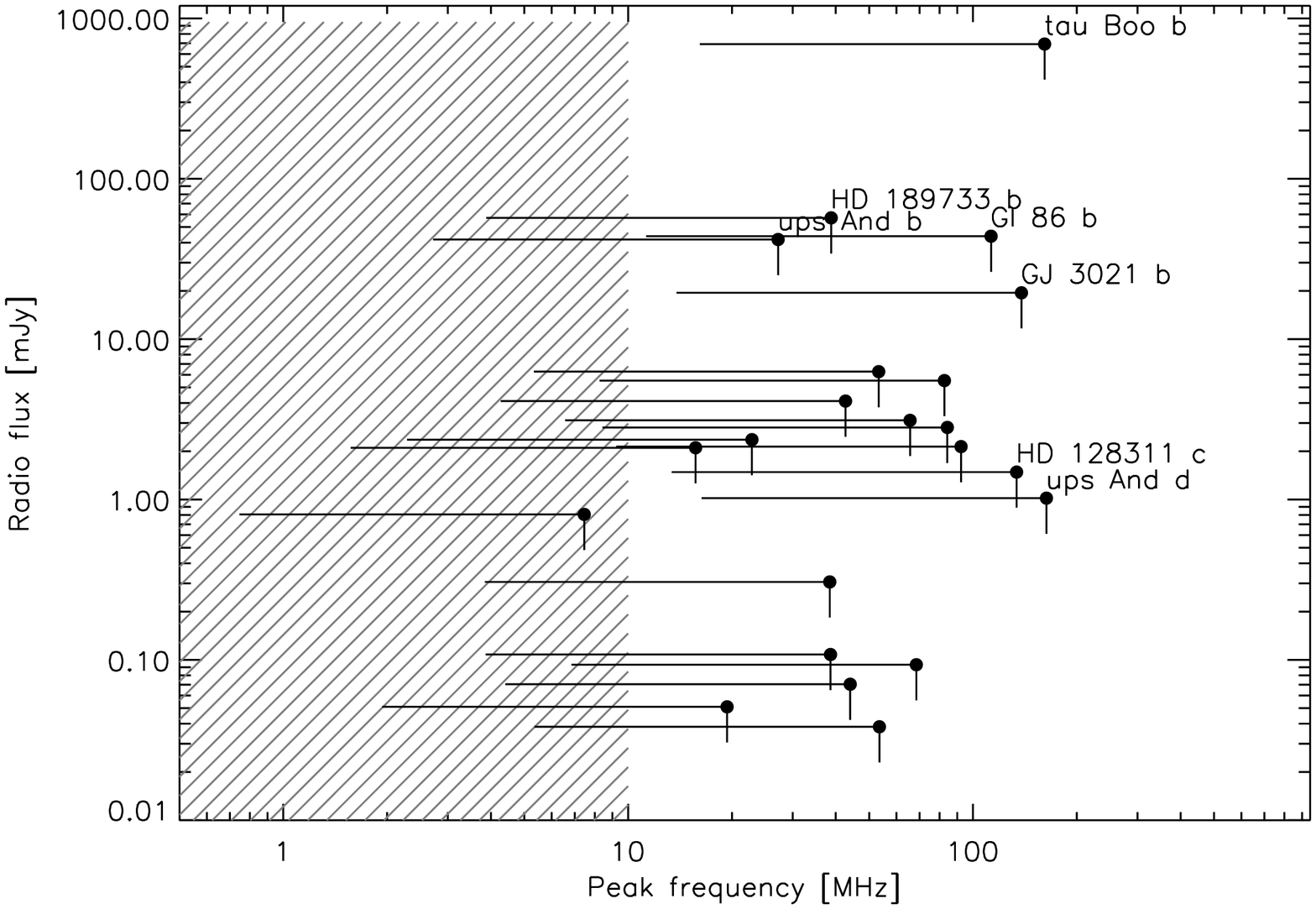}}
  \caption{\label{fig:radiopeak}Radio flux for the extrasolar planets
    of Table\,\ref{tab:stars}. For each planet, the expected radio
    flux at Earth is plotted for the frequency range $(0.1-1)f_{\rm
      ce}$, which is the expected range of strong emission as in the
    case of Jupiter. The hatched area marks the frequency range below
    the ionospheric cutoff.}
\end{figure*}

We have calculated the radio flux and cutoff emission frequency for
the known planets of stars within 20\,pc and with X-ray detections.
Planet parameters are from \emph{The Extrasolar Planets
  Encyclopaedia}. The results are given in Table\,\ref{tab:stars} and
are plotted in Fig.\,\ref{fig:radiopeak}. \citet{Stevens05} assumes
the peak radio flux occurs at the electron cyclotron frequency of the
equatorial surface field, i.e. at half the cutoff frequency.  We mark
the frequency range, $(0.1-1)f_{ce}$, over which significant radio
emission can be expected for each planet by horizontal lines in
Fig.\,\ref{fig:radiopeak}. The extension of the range to lower
frequencies is rather arbitrary but indicative for the emission from
most of the planets of our solar system \citep[e.g.,][]{Zarka92}.  We
include only planets that are more massive than about 0.5\,M$_{\rm
  Jup}$ because in Saturn-sized or smaller planets helium separation
may lead to stable stratification at the top of the electrical
conducting region \citep{Stevenson80}. The associated reduction of the
surface field strength is difficult to quantify.

\begin{table*}
  \center
  \caption{\label{tab:stars}Parameters of known planets around stars
    with X-ray detections within 20\,pc}
  \begin{tabular}{lccrrccrr}
    \hline
    \hline
    \noalign{\smallskip}
    Planet Name & Planet mass & $a$ & d$^1$ & $\dot{M}$ & Age & $B_{\rm dip}^{\rm pol}$ & radio flux$^1$ & $f_{\rm ce}$\\
    & [${\rm M_{Jup}}\sin{i}$] & [AU] & [pc] & [$\dot{M_\odot}$] & [Gyr] & [G] & [mJy] & [MHz]\\
    \noalign{\smallskip}
    \hline
    \noalign{\smallskip}
    \input{Planets_table.tex}
    \noalign{\smallskip}
    \hline
  \end{tabular}\\
  $^1$The distance to Jupiter was set to 5.2\,AU for the calculation of its radio flux.
\end{table*}

The maximum radio flux predicted for known extrasolar planets is about
700 mJy in the case of $\tau$~Boo\,b. Maximum emission frequencies are
between 7 and 160\,MHz, i.e., in most cases above the ionospheric
cutoff frequency of 10\,MHz. However, when the maximum frequency is
less than 20\,MHz the peak radio emission may fall below the
ionospheric cutoff. The predicted flux for planets other than
$\tau$~Boo\,b is at least an order of magnitude smaller.  The fluxes
for $\upsilon$~And\,b, Gl~86\,b, and HD~189733\,b fall into the range
of 40-60 mJy, and their maximum frequencies are well above the
ionospheric cutoff. For GJ~3012\,b we predict 20\,mJy, and all other
planets fall below 10~mJy.

Our predicted radio flux values are similar to those in
\citet{Stevens05}, which is mainly due to the fact that the distance
to the object is an important factor in the observed radio emission,
and that we use the same assumptions for the stellar mass loss rate.
Our Fig.\,\ref{fig:radiopeak} can be compared to Figs.\,1--3 in
\citet{Griessmeier07a}. In general, the range of radio frequencies and
radio flux are comparable but can differ substantially between
individual objects. Note that we only show estimates for the planets
within 20\,pc. These are the most likely candidates for the detection
of radio emissions.

\section{Summary and Discussion}

We applied the energy flux scaling relation from \citet{Christensen09}
to estimate the magnetic field evolution on giant extrasolar planets
and brown dwarfs. This magnetic field scaling is independent of the
rotation of the objects given that they rotate above a critical
rotation limit, which probably is the case for isolated brown dwarfs,
for young exoplanets, and for exoplanets in orbits not too close to
their central star. Close-in planets suffer tidal braking. This
applies to all candidate planets for which we predict a radio flux
above 10\,mJy, except GJ~3021\,b.  However, for planets that are very
close to their host star the synchronous rotation rate may still lie
above the critical limit.  The critical period is on the order of four
days in M-dwarfs, and if we assume the that the critical limit is
somewhat higher, the top candidate for the detection of radio
emissions, $\tau$~Boo\,b (3.3\,d), as well as $\upsilon$~And\,b
(4.6\,d) and HD~189733\,b (2.2\,d) rotate rapidly enough. At this
point, we cannot say more about the real critical limit. Gl~86\,b
(15.8 d) is probably rotating too slowly and our magnetic field and
radio flux estimates are likely too high.

Because energy flux scales with luminosity, young exoplanets have
magnetic fields about an order of magnitude higher than old
exoplanets. Brown dwarfs go through a similar evolution but may go
through a temporal magnetic field maximum depending on the details of
the luminosity and radius evolution. Very low-mass stars build up
their magnetic fields during the first few $10^6$ years and maintain a
constant magnetic field during their entire lifetime as long as they
are not efficiently braked. The latter is consistent with observations
of Zeeman splitting in FeH lines, which directly yields a measurements
of the average magnetic field \citep{Reiners07}.

So far, no Zeeman measurements were successful in old brown dwarfs. On
the other hand, \citet[][ and references therein]{Hallinan08} confirms
the observation of electron cyclotron masers on three ultra-cool
dwarfs that are probably brown dwarfs older than a few hundred Myr.
This means, these ultra-cool dwarfs emit radio emission by a mechanism
similar to the one discussed for planets in this paper. In particular,
from radio detections at 4.88\,GHz and 8.44\,GHz, Hallinan et al.
conclude that these brown dwarfs have regions of magnetic fields with
$B>1.7$\,kG and $B>3$\,kG, respectively. All three objects are
high-mass brown dwarfs (0.06 -- 0.08\,M$_{\odot}$) older than a few
hundred Million years. The (average) magnetic field prediction for
this class of objects is $\sim 3$\,kG from our models, which is in
good agreement with the observations. Another prediction from our
model is that old brown dwarfs at several Gyr age have weaker fields.
This needs to be tested in other (older) objects, which requires good
estimates of the ages of (old) brown dwarfs.  Furthermore,
measurements at other radio frequencies are desirable in order to
better constraint their maximum field strength.

\citet{Reiners09b} failed to detect kG-strength magnetic fields in
four young ($\la 10$\,Myr) \emph{accreting} brown dwarfs that are
rapidly rotating ($v\,\sin{i} > 5$\,km\,s$^{-1}$). According to our
scenario, the objects should have average magnetic fields of 1--2\,kG,
but upper limits of about one kG are found in all of them. This may be
a hint for a deviation from our scaling law suggesting that brown
dwarfs do not follow this rule, at least not at ages below
$\sim10$\,Myr. Another alternative is that (at least in brown dwarfs)
the average magnetic field may be altered by the presence of an
accretion disk \citep[depending on the position of the X-point and the
amount of trapped flux; see ][]{Mohanty08}.

Radio flux predictions for giant exoplanets based on our scaling
relation are not dramatically different from earlier estimates by
\citet{Farrell99}, \citet{Stevens05}, and
\citet{Griessmeier07a}. Still, the uncertainties in the predicted
radio flux are much larger than the actual values. First, our
assumption of a homogenous magnetic field interacting with an
isotropic wind is certainly oversimplified. Quantitative uncertainties
of our flux estimates mainly come from uncertainties in the dipole
moment (we estimate a factor of 3) and in the mass-loss rate (another
factor of 3). Individual values of the radio flux are therefore
uncertain by at least a factor of 5, but differential comparison
between the radio flux predictions are likely to be more trustworthy
because the objects are in general rather comparable.

We estimate that only few extrasolar planets may emit radio flux
larger than 10\,mJy; $\tau$~Boo\,b is the strongest potential emitter
with $P \sim 700$\,mJy. The maximum emission frequencies are well
above the ionospheric cutoff frequency of 10\,MHz in most cases so
that detections of radio emission should not be hampered by the
ionosphere.

The uncertainties of magnetic field and radio emission predictions are
still very large. Nevertheless, several different scaling scenarios
now exist and the one of \citet{Christensen09} was shown to
successfully reproduce magnetic fields of both planets and stars. The
detection of magnetic fields in giant extrasolar planets through other
techniques like the Zeeman effect will probably remain out of reach
for a few decades, at least in old planets with fields that are
probably well below 100\,G.  However, very young and massive planets
may harbor magnetic fields up to one kilo-Gauss, which may become
detectable in a few systems in the near future.

Radio emission from extrasolar planets has not been detected so far,
but the technology for a detection may become soon available with
facilities like LOFAR and others. Our scaling relation favors a
different set of planets than that suggested by \citet{Griessmeier07a}
although $\tau$~Boo\,b remains the most obvious choice. Our scaling
relation can easily be applied to extrasolar planets that will be
detected in the future, and target selection for radio observation
campaigns can be based on these predictions.

\begin{acknowledgements}
  A.R. acknowledges research funding from the DFG as an Emmy Noether
  fellow (RE 1664/4-1).
\end{acknowledgements}

\end{document}

%% file: Planets_table.tex
        Jupiter &  1.00 &  5.20 &     &   1.0 &   4.5 &    9 & 4.E+09 &  27 \\
\noalign{\smallskip}
  eps Eridani b &  1.55 &  3.39 &   3.2 &  25.9 &  1.7 &   19 &   6.3 &  53 \\
   Gliese 876 b &  1.93 &  0.21 &   4.7 &   0.1 &  2.4 &   23 &   3.1 &  66 \\
   Gliese 876 c &  0.56 &  0.13 &   4.7 &   0.1 &  2.4 &    6 &   2.1 &  16 \\
       GJ 832 b &  0.64 &  3.40 &   4.9 &   0.2 &  2.0 &    7 &   0.1 &  19 \\
     HD 62509 b &  2.90 &  1.69 &  10.3 &   0.3 &  5.6 &   24 &   0.1 &  68 \\
        Gl 86 b &  4.01 &  0.11 &  11.0 &   9.4 &  2.9 &   40 &  43.8 & 113 \\
    HD 147513 b &  1.00 &  1.26 &  12.9 & 150.4 &  0.8 &   15 &   4.1 &  43 \\
      ups And b &  0.69 &  0.06 &  13.5 &  20.2 &  1.4 &   10 &  41.8 &  27 \\
      ups And c &  1.98 &  0.83 &  13.5 &  20.2 &  1.4 &   30 &   2.8 &  84 \\
      ups And d &  3.95 &  2.51 &  13.5 &  20.2 &  1.4 &   58 &   1.0 & 163 \\
 gamma Cephei b &  1.60 &  2.04 &  13.8 &   1.1 &  3.6 &   16 &   0.1 &  44 \\
       51 Peg b &  0.47 &  0.05 &  14.7 &   0.2 &  6.2 &    3 &   0.8 &   7 \\
      tau Boo b &  3.90 &  0.05 &  15.0 & 198.5 &  0.8 &   58 & 692.2 & 161 \\
       HR 810 b &  1.94 &  0.91 &  15.5 & 103.9 &  0.8 &   30 &   5.5 &  83 \\
    HD 128311 b &  2.18 &  1.10 &  16.6 &  39.9 &  0.9 &   33 &   2.1 &  92 \\
    HD 128311 c &  3.21 &  1.76 &  16.6 &  39.9 &  0.9 &   48 &   1.5 & 134 \\
     HD 10647 b &  0.91 &  2.10 &  17.3 &  22.9 &  1.4 &   14 &   0.3 &  38 \\
      GJ 3021 b &  3.32 &  0.49 &  17.6 & 170.2 &  0.8 &   49 &  19.5 & 138 \\
     HD 27442 b &  1.28 &  1.18 &  18.1 &   1.9 &  2.7 &   14 &   0.1 &  39 \\
     HD 87883 b &  1.78 &  3.60 &  18.1 &   2.6 &  3.3 &   19 &   0.0 &  54 \\
    HD 189733 b &  1.13 &  0.03 &  19.3 &  17.3 &  1.7 &   14 &  57.1 &  39 \\
    HD 192263 b &  0.72 &  0.15 &  19.9 &   7.1 &  2.5 &    8 &   2.4 &  23 \\